%% file: main.tex
\documentclass[sigconf]{acmart}
\usepackage[a-1b]{pdfx}
\usepackage[utf8]{inputenc}
\usepackage{amsmath}
\usepackage{graphicx}
\usepackage{booktabs}
\usepackage{pgfplots}
\usepackage{multirow}

\usepackage[symbol]{footmisc}
\copyrightyear{2021}
\acmYear{2021}
\setcopyright{acmcopyright}\acmConference[CIKM '21]{Proceedings of the 30th ACM
International Conference on Information and Knowledge Management}{November
1--5, 2021}{Virtual Event, QLD, Australia}
\acmBooktitle{Proceedings of the 30th ACM International Conference on Information
and Knowledge Management (CIKM '21), November 1--5, 2021, Virtual Event, QLD,
Australia}
\acmPrice{15.00}
\acmDOI{10.1145/3459637.3482101}
\acmISBN{978-1-4503-8446-9/21/11}
\title{Query-driven Segment Selection for Ranking Long Documents}

\author{Youngwoo Kim, Razieh Rahimi, Hamed Bonab and James Allan}

\affiliation{%
  \institution{University of Massachusetts Amherst}
  \city{Amherst} 
  \state{MA}
  \country{USA}
}
\email{{youngwookim, rahimi, bonab, allan}@cs.umass.edu}

\date{}
\input{0_abstract}
\DeclareMathOperator*{\argmax}{arg\,max}
\DeclareMathOperator*{\argmin}{arg\,min}


\begin{document}
\fancyhead{}
\newcommand{\ProposedMethod}{{BeST}}

\newcommand{\James}[1]{{\bf \color{red} [James: ``#1'']}}
\newcommand{\Youngwoo}[1]{{\bf \color{green} [Youngwoo: ``#1'']}}
\newcommand{\Rab}[1]{{\bf \color{blue} [Rab: ``#1'']}}
\newcommand{\Todo}[1]{{\bf \color{red} [TODO: ``#1'']}}
\newcommand{\Negin}[1]{{\bf \color{violet} [Negin: ``#1'']}}

\maketitle
\input{1_Introduction}
\input{2_related_work}
 \input{3_method}
\input{4_experiments}
\input{9_conclusion}

\input{10_acknowledgment}

\bibliographystyle{ACM-Reference-Format}
\bibliography{bib}
\end{document}

%% file: 0_abstract.tex
\begin{abstract}
Transformer-based rankers have shown state-of-the-art performance. However, their self-attention operation is mostly unable to process long sequences. One of the common approaches to train these rankers is to  heuristically select some segments of each document, such as the first segment, as training data. However, these segments may not contain the query-related parts of documents. To address this problem, we propose  query-driven segment selection from long documents to build training data. The segment selector provides relevant samples with more accurate labels and non-relevant samples which are harder to be predicted. The experimental results show that the basic BERT-based ranker trained with the proposed segment selector significantly outperforms that trained by the heuristically selected segments, and performs equally to the state-of-the-art model with localized self-attention that can process longer input sequences. Our findings open up new direction to design efficient transformer-based rankers.
\end{abstract}

%% file: 1_Introduction.tex
\section{Introduction}
Transformer-based rankers have shown state-of-the-art performance for many tasks. However the self-attention operation in these models cannot be applied to long sequences, as its complexity scales quadratically with the sequence length~\cite{vaswani2017attention}. Unfortunately, documents in ad-hoc information retrieval are usually much longer than the length that the self-attention operation can be applied to. We propose a query-driven segment selector approach to effectively train the basic BERT-based ranker~\cite{dai2019deeper} for long documents.
 
There are two common approaches for handling long documents in transformer models. The first addresses the problem by dividing the document tokens into segments of similar length and  applying the self-attention mechanism only within those segments (local self-attention) then aggregating vectors from segments to get the final score.~\cite{beltagy2020longformer, li2020parade,jiang2020long, macavaney2019cedr}. A second approach also starts by dividing the document tokens into segments and treating each of them as if it were a document in its own right. If the segments are small enough, the complexity problem is similarly limited. The starting document's score is typically found with an aggregation (e.g., average or max) of the segments' scores~\cite{yilmaz2019applying, dai2019deeper}. It appears that the second approach is more commonly used. Indeed, we note that the first approach requires limiting the size of the input document in the training~\cite{jiang2020long} (though allows it to be larger).



To train the second category of approaches, we can consider different segments of a long document to use for training samples. For example, we can use all segments, a randomly selected subset, the first or last segment. Choosing the first segment is very common -- and very appropriate for news stories that often provide a summary of the key points in the first few paragraphs. However the segments are chosen, each provides a separate training instance. The relevance label of each selected segment of a document is inherited from the relevance label of the entire document.

In ad-hoc information retrieval, a document is considered relevant to a query even if only a small part of the document is relevant to the query~\cite{TREC}. This relevant information can occur in any part of a document and could be as short as a single sentence or even a clause. 
Thus, it is possible that heuristically selected segments do not contain the part of document that is relevant to the query and, indeed, may actually be completely non-relevant. We demonstrate in this study that such noise in the training data leads to poor training of the model and harms performance. 

Instead of heuristic selection of document segments to train a Transformer-based ranker, we propose to explicitly select which segment of a  document should be used for the training. Our proposed method, which we named \ProposedMethod\ (BEst Segment Training), scores different segments of a document and selects the one that is most related to the query to build the training data for the subsequent ranker.
\ProposedMethod\ not only provides more accurate labels for segments from relevant documents, but also chooses the segments from \emph{non-relevant} documents with the highest similarity to the query and builds hard training samples which have been shown to  impact the effectiveness of learned rankers.

We apply \ProposedMethod\ to the basic BERT-based ranker~\cite{dai2019deeper} and compare it with the baselines that only use first segments of documents as training data, and other Transformer-based rankers that are designed to handle longer sequence (local self-attention)~\cite{hofstatter2020local,jiang2020long}.
The experiments on the TREC 2019 Deep Learning Track and Robust04 datasets demonstrate that \ProposedMethod\ significantly improves the effectiveness of the ranker. 
This finding presents new opportunities for the design and development of efficient Transformer-based rankers. 

%% file: 2_related_work.tex
\vspace{-0.5 cm}
\section{Related work}

To get a document level score with BERT, applying BERT to applicable length segments and aggregating scores is the most widely used strategy~\cite{dai2019deeper, yilmaz2019cross}.
For the models where the score is based on cosine similarity of embeddings between a query and documents, the maximum cosine similarity scores could be pooled and fed to final score generating layers~\cite{macavaney2019cedr, khattab2020colbert, hofstatter2020local}. While these embedding similarity based approach has benefit of inference in document scoring, they still have the problem that the training data are heuristically selected segments from the documents.

Some other works proposed a document level modeling, which either uses much longer sequence with efficient attention or a hierarchical structure \cite{hofstatter2020local,jiang2020long}.
TKL transformer encodes long document by encoding a number of overlapping windows separately~\cite{hofstatter2020local}. The authors show that including longer sequences for neural ranking models help to retrieve longer documents and this results in increased retrieval performance \cite{hofstatter2020local}.
PARADE model divides a document into a number of segments and encodes each of them using Transformer. Encoded segments are again fed to another Transformer to get the final document level score~\cite{li2020parade}~\footnote{Results of the PARADE model is not directly comparable because they used transfer learning setup from MSMARCO passage ranking to Robust04}.
QDS-Transformer encodes the long texts with fixed patterns of attentions which allow local attention among neighboring content tokens, and combines them with long distance attention~\cite{jiang2020long}. 
A few revised Transformer architectures were proposed to solve the quadratic cost problem. Based on the intuition that not all tokens have to directly attend to all the other tokens, they limit the number of tokens that self-attentions could be applied~\cite{tay2020efficient}.
Self-attention are either limited to a fixed pattern tokens~\cite{beltagy2020longformer} or learnable patterns~\cite{wang2020linformer} such as using locality sensitive hashing~\cite{kitaev2020reformer}.

%% file: 3_method.tex
\section{ Document Segment Selection}
\label{sec:method}

At a high level, the goal of our method is to identify the most related segments of documents to queries and use those segments in the training. The selection of the most related segments can be done by another ranker, which can have the same the architecture as the ranker to be trained. In this sense, the training of the ranker and the segment selection can be iterated to even improve the performance. 

We assume a set of training instances $\mathcal{S}$ where each instance consists of a query and a set of relevant and non-relevant documents $(q, D_q^{+,-})$. The goal is to train a ranker that ranks a set of documents with respect to a query where its scoring function $f_\theta(.)$ imposes a token length limit on the input sequence. 

Let $d_i$  denote the $i$\textsuperscript{th}  segment of  document $d$ and $l$ be a pairwise loss function, which will be $l(y_1, y_2)=max(0, 1 - y_1 + y_2)$ in the case of the pairwise hinge loss. 
The loss function for the first segment-based training can be expressed as
\begin{equation}
        \mathcal{L}(\theta) = \sum_{q \in \mathcal{S}} \sum_{d^+, d^- \in D_q^{+,-}} l( f_{\theta}(q, d^+_0), f_{\theta}(q, d^-_0)),~\label{eq:first}
    \vspace{-1mm}
\end{equation}
where $\theta$ is the set of learnable parameters of the scoring function $f$ and $d^+_0$ and $d^-_0$ denote the first segments of a relevant and non-relevant document to query $q$, respectively. Typically, $\theta$ is initialized with pre-trained language models, such as BERT.

We propose to select the document segments  most relevant to queries for building effective training data, rather than merely using the first segments of documents.
Let $i_{q,d}$ denote the index of the segment in document $d$ that is most related to the query $q$.
The set ${\mathcal{I} =\{i_{q,d}| (q, d) \in \mathcal{S}\}}$ then denotes the set of segment indices for all query-document pairs in $\mathcal{S}$. 
We propose to learn the scoring function $f_{\theta}(.)$ according to  $\mathcal{I}$, which is our best guess of the segment indices  most related to queries.
In contrast to Equation~\ref{eq:first} where only the first segments are considered, the loss function for training with the segment selection can then be expressed as 
\begin{equation}
   \mathcal{L}(\theta, \mathcal{I}) = \sum_{q \in \mathcal{S}} \sum_{d^+, d^- \in D_q^{+,-}} l(f_{\theta}(q, d^+_j), f_{\theta}(q, d^-_k))~\label{eq:loss_common}
\end{equation}
where $j$ and $k$ are given as $i_{q,d^+}$ and $i_{q,d^-}$.
Estimating $\mathcal{I}$ and updating $\theta$ based on it can be considered as training a ranking model with selected segments from documents. 
Note that both $\theta$ and $\mathcal{I}$ are unknown at the beginning and need to be estimated. 
To estimate $\theta$ and $\mathcal{I}$, we use an iterative approach that includes complete and incomplete data, which is similar to the EM algorithm. 
For example, once we train $\theta^{(1)}$,
each segments can be scored with $f_{\theta^{(1)}}$, and $\mathcal{I}^{(2)}$ could be obtained by taking maximum of them. Then, $\mathcal{I}^{(2)}$ is used to train $\theta^{(2)}$, and so on.

In the \textbf{first iteration}, the goal is to estimate $\mathcal{I}^{(1)}$ and $\theta^{(1)}$. 
To estimate $\mathcal{I}^{(1)}$, we introduce $\theta^{(0)}$, from which $\mathcal{I}^{(1)}$ will be calculated.
The iteration begins with initializing both $\theta^{(0)}$ and $\theta^{(1)}$ with pre-trained parameters (e.g., BERT) or random assignments.
$\theta^{(0)}$ is trained using all segments of both relevant and non-relevant documents. The loss function for $\theta^{(0)}$ can be expressed as 
\begin{equation}
   \mathcal{L}^{(0)}(\theta^{(0)}) = \sum_{q \in \mathcal{S}} \sum_{d^+, d^- \in D_q^{+,-}} \sum_{j=0}^{k-1} l( f_{\theta^{(0)}}(q, d^+_j), f_{\theta^{(0)}}(q, d^-_j)), \nonumber
\end{equation}
where $j$ is the segment index and $k$ is the maximum number of segments to be considered per document. 
While $\theta^{(0)}$ is being updated, $\mathcal{I}^{(1)}$ is calculated as indices of maximum scoring segments by $f_{\theta^{(0)}}$:
\begin{equation}
    \mathcal{I}^{(1)} = \{i_{q,d} | i_{q,d} = \argmax_i f_{\theta^{(0)}}(q, d_i), (q,d) \in \mathcal{S}\}.
\end{equation}
Based on $\mathcal{I}^{(1)}$, $ \theta^{(1)}$ is optimized using the loss function shown in equation~\ref{eq:loss_common}, which can be written as $\theta^{(1)} = \argmin_{\theta} \mathcal{L}(\theta, \mathcal{I}^{(1)}).$

Exceptionally for the first iteration we optimize $\theta^{(1)}$ at the same time that $\theta^{(0)}$ is optimized. 
After the first iteration, it is possible to take scorer $f_\theta^{(1)}$ as the final scorer as it outperforms the first-segment trained models. However, additional iterations of segment-selection and training could bring even further improvements. 

From the \textbf{second iteration onward}, the previously trained model $f_{\theta^{(.)}}$ calculates $\mathcal{I}$ by scoring each of the document segments and selecting the one with the highest score. Specifically, at iteration $n+1$, $\mathcal{I}^{(n+1)}$ is calculated as
\begin{equation}
    \mathcal{I}^{(n+1)} = \{i_{q,d} | i_{q,d} = \argmax_i f_{\theta^{(n)}}(q, d_i), (q,d) \in \mathcal{S}\}.
\end{equation}
The loss function is given as
\begin{align}
    \mathcal{L}(\theta^{(n+1)}, \mathcal{I}^{(n+1)}) = \sum_{q \in \mathcal{S}} \sum_{d^+, d^- \in D_q^{+,-}} l(f_{\theta^{(n+1)}}(q, d^+_j), f_{\theta^{(n+1)}}(q, d^-_k)). \nonumber
\end{align}
where $j$ and $k$ are given as $i^{(n)}_{q, d^+}$ and $i^{(n)}_{q, d^-}$.

Parameters $\theta^{(n+1)}$ are then updated to minimize $\mathcal{L}(\theta^{(n+1)}, \mathcal{I}^{(n+1)})$.
Note that $\theta^{(n+1)}$ is newly initialized instead of updating $\theta^{(n)}$ to minimize the effect of non-optimal training data.

We would like to clarify that each iteration is a complete training procedure which includes (possibly) multiple epochs of training, early-stopping, and model selections. 
In practice, estimated $\theta^{(n)}, \mathcal{I}^{(n)}$ from later iterations may not be better than those from earlier iterations, 
probably because the model has many parameters and prone to over-fitting. Thus, the iterations need to be early-stopped based on the validation performance. 

%% file: 4_experiments.tex
\section{Experiment Design}


\textbf{Datasets.}
We conduct our experiments on two document ranking datasets, Robust04 and TREC 2019 DL (Deep Learning Track).
TREC 2019 DL is a large dataset for adhoc ranking~\cite{trec19dl}.
Its training data contains about 384,000 judged query and document pairs which are built from MS MARCO passage ranking datasets~\cite{msmarco}, by taking any document that contains a relevant passage (from MS MARCO passage ranking datasets) as a relevant document. The test split contains 43 queries, whose top candidates are exclusively judged.
Robust04 is a collection of news articles containing 249 topics with relevance judgment.
Robust04 topics have two types of queries, title queries and description queries, and we evaluate on both of them.

The documents in TREC 2019 DL has median length is 797 tokens, with the 90\textsuperscript{th} percentile including 3,259 tokens (in subword).
From Robust04 collection, the median length is 594 tokens, with the 90\textsuperscript{th} percentile including 1,561 tokens.

We applied \ProposedMethod\ to the BERT-based ranker~\cite{dai2019deeper} which fine-tunes BERT with pooling the [CLS] token's representation through a fully connected layer. To control the performance differences by other factors, we applied the same BERT-based ranker for all the baselines and \ProposedMethod. All the BERT models in our experiments used BERT-Base uncased, as opposed to BERT-Large or other variants. 

\input{Table/exp_all2}

\textbf{Compared models.}
We implemented three baselines models: BERT (First P), BERT (Max P), and BERT (Gold P).
Both BERT (First P) and BERT (Max P) are trained by taking the first segment of each document. At inference time, BERT (First P) only uses the first passage while BERT (Max P) uses the maximum score of all passages. 
BERT (Gold P) fine-tunes BERT by using passages of documents that are relevant to the query according to the MS MARCO \emph{passage} ranking dataset. This is possible because the TREC 2019 DL dataset for \emph{document} ranking is built based on the MS MARCO \emph{passage} ranking dataset, thus the relevant passages of documents can be obtained. After aligning relevant passages to documents dataset by query IDs and contents, 58\% of relevant passages are used from MS MARCO passage ranking.
We created this baseline because the goal of \ProposedMethod\  is to use the most relevant segments of documents. Thus,  BERT (Gold P) is an oracle that provides an upper bound to the performance based on ground-truth data.

We also report the results of IDST which is one of the best performing runs in the original TREC 2019 Deep Leaning TRACK~\cite{trec19dl}. 
There are other runs which have better performance than IDST, but they are inappropriate for comparison because they either use full-ranking instead of re-ranking, use larger pre-trained models, or use ensemble models. Still, IDST has a few other factors that contribute to its high performance, such as additional language model pre-training and using the entire passage-level labels as training data~\cite{yan2019idst}. 
We also consider  existing models that are especially designed to handle long inputs. 
QDS-Transformer is a ranker with local and hierarchical attention which can efficiently model long sequences~\cite{jiang2020long}. RoBERTa (Max) and Longerformer-QA were used as the baselines for QDS-Transformer. 
TKL processes long sequences by local self-attention mechanism~\cite{hofstatter2020local}. Unlike other models, TKL uses Glove embeddings instead of pre-trained language models. 
Unless mentioned otherwise, all the methods use a fixed number of leading tokens (which can vary from 512 to 4000 depending on models) for the document representation in the training.


\textbf{Implementation details.} For TREC 2019 DL, we used pair-wise hinge loss by taking one positive document (judged) and one negative from the top 100 documents (not judged). Documents are split into segments only at the sentence boundary. Instead of splitting documents into segments of the maximum possible  length (which is determined by the input size limit), we randomized the length of  each segment for  training, so that the model does not predict  document relevance based on the segment length.  The document title is repeated in each of its segments. 

For Robust04, we used point-wise cross-entropy loss by taking one positive and 10 negative documents for each query. Because Robust04 has a smaller number of queries than TREC 2019 DL, using pairwise loss makes more repetition of positive
documents during the training, which was less effective than using diverse negative documents in the training. 
For \ProposedMethod\, we only used up to four leading  segments (less than 2,048 tokens) of the documents in the training, due to the computational efficiency. At inference time, however, all document segments are used. Document segments are  obtained by a non-overlapping moving window through contents.



We compare all models in the re-ranking setup, where only the top-100 documents retrieved for each query are used for training and inference of neural rankers.  
The Robust04 dataset does not have designated training and test splits, therefore we repeat each experiment 5 times using 5-fold cross validation similar to previous work~\cite{dai2019deeper, jiang2020long}. 
To compare our experimental results with those reported in existing work, we report the commonly used metrics for each of dataset, which are NDCG@10 for TREC 2019 DL and NDCG@20 for Robust04. We also list mean reciprocal rank (MRR) for the development split of TREC 2019 DL, which is used for validation including early-stopping of training. 



\textbf{Results.}
Table~\ref{table:exp_main} shows the performance of the ranker trained with \ProposedMethod\ and other models. 
The ranker trained with \ProposedMethod\ outperforms those trained with the first segments as supervision data, BERT~(First P) and BERT~(Max P). The differences between the models trained with first segments and \ProposedMethod\ are mostly statistically significant ($p < 0.01$) using paired t-test.
This result strongly demonstrates the benefit of segment selection to remove noise in training data. It is noteworthy that \ProposedMethod\ achieves a similar performance as BERT~(Gold P), which is trained with ground-truth passage level evidences. The reason could be that the segment selection of \ProposedMethod\ is as precise as the human judged passage-level evidences, or \ProposedMethod\ has other benefits such as more difficult negative examples. 
In the TREC 2019 DL dataset, \ProposedMethod\ shows a similar performance to QDS-Transformer~\cite{jiang2020long}, and outperforms TKL~\cite{hofstatter2020local}. TKL being worse than both \ProposedMethod\ and QDS-Transformer could be attributed to only using Glove and not using bigger pre-trained language models. 

In the Robust04 dataset, RoBERTa~(Max P) and QDS-Transformer under-perform BERT~(Max P) and \ProposedMethod. We believe the implementation of RoBERTa~(Max P) and QDS-Transformer~\cite{jiang2020long} are less-optimal in terms of negative sampling strategy or choices of loss functions, as our implementation of BERT~(Max P) matches the number reported in another previous work~\cite{dai2019deeper}. Nevertheless, \ProposedMethod\ is showing consistent improvement over BERT~(Max P) on both title and description queries of the Robust04 dataset.



Theoretically, models that process long sequences of 2,048 or more tokens at once, such as  QDS-Transformer and TKL, have some advantages over \ProposedMethod. First, they have the potential of identifying relevance evidences that are far apart of each other in a document and thus appear in different document segments.  
Second, with the same amount of tokens to be used as training data for rankers, these models have the potential of processing more tokens from document bodies. This is because these models build one training instance from each document, while building multiple training instances from different document segments requires query tokens and document titles to be repeated. 
Despite their advantages, QDS-Transformer and others only consider the leading tokens of documents for the inferences, which can be problematic for documents that are longer than their input length limit. 
The performance gap between \ProposedMethod\ and IDST could be attributed to using a larger amount of training data (we used 45\% of the passage-ranking dataset that can be aligned with documents in the TREC 2019 DL dataset) and to an additional pre-training strategy specifically designed for the task~\cite{yan2019idst}, which is shown to be effective in other tasks as well~\cite{Lan2020ALBERT, Wang2020StructBERT}.


\input{Table/exp_seg_sel2}
In addition to extrinsic evaluation of  \ProposedMethod\ through ranking performance, we also  intrinsically evaluate the performance of \ProposedMethod\ in selecting document segments. For this purpose, we consider a selected document segment with respect to a query as  correct  if the selected segment contains the relevant passage for the corresponding query in the ground-truth of the MS MARCO passage-ranking dataset. 
\ProposedMethod\ and baselines rank the segments of each document, and their performance is measured using Precision at the top 1 segment. We report the performance of \ProposedMethod\ at different iterations corresponding to different $f_{\theta^{(n)}}$ in Section~\ref{sec:method}. As baselines for the intrinsic evaluation, we report the performance of Random ranking of document segments and ranking by BERT~(Max P). Table~\ref{tab:seg_sel} shows the performance of ranking document segments with respect to queries. For reference, the performance of the document ranking task over the dev split is also listed in the table. The results show that \ProposedMethod\ consistently outperforms the baselines in all iterations, and the performance of \ProposedMethod\ increases as the iteration number increases. 
Another observation is that the performance of segment selection and document ranking are closely correlated. Using the performance of ranking over the dev split to stop training of  segment selection, the learned model at iteration 3 will be selected as it achieves the highest MRR. 

%% file: Table/exp_all2.tex
\begin{table*}[t]
\small
\caption{Ranking performance on the TREC 2019 DL and Robust04 datasets. The metrics for each dataset are the ones that are widely used for the datasets. Input size is the number of (word or subword) tokens that the models take for each instance. The superscripts on the scores of  \ProposedMethod\ indicates the models that the difference is statistically significant ($p<0.01$). }
\label{table:exp_main}
\begin{tabular}{l|ll|cc|cc|ccc}
\toprule
    \multicolumn{3}{c|}{}                              & \multicolumn{2}{c|}{TREC 2019 DL}             & \multicolumn{2}{c|}{Robust04}          &                            &                     &                     \\ \cline{4-10}
     \multicolumn{3}{c|}{}                             & \textbf{Dev} & \textbf{Test} & \textbf{Title} & \textbf{Description} & \textbf{Pre-trained} & \multirow{2}{*}{\textbf{Input size}} & \multirow{2}{*}{\textbf{From}}       \\ \cline{4-7}
\multicolumn{3}{c|}{} & \textbf{MRR} & \textbf{NDCG@10} & \textbf{NDCG@20} & \textbf{NDCG@20} & \textbf{model} \\ \hline 

\multirow{3}{*}{Baselines}  & 1 & BERT (First P)                    & 0.375              & 0.637                   & 0.464          & 0.505                & BERT                       & 512                 &                     \\
 & 2 & BERT (Max P)                      & 0.349              & 0.626                   & 0.471          & 0.519                & BERT                       & 512                 &                     \\
 & 3 & BERT (Gold P)                     & 0.363              & 0.664                   & -              & -                    & BERT                       & 512                 &                     \\\hline 
Segment selection & 4 & \ProposedMethod                    & 0.388              &  0.664$^2$                   &     \textbf{0.487}$^{12}$  & \textbf{0.537}$^{12}$                & BERT                       & 512                 &                     \\\hline 
\multirow{2}{*}{Models w/ short input}& 5 & IDST                              & -                  & \textbf{0.704}                   & -              & -                    & BERT+@                     & 512                 & \cite{yan2019idst}         \\
 & 6 & RoBERTa (Max P)                     & 0.320               & 0.630                    & 0.439          & -                    & RoBERTa                    & 512                 & \cite{jiang2020long}       \\ \hline 
\multirow{3}{*}{Models w/ long input} & 7 & Longformer-QA                     & 0.326              & 0.627                   & 0.448          & -                    & Longformer                 & 2,048                & \cite{jiang2020long}       \\
 & 8 & QDS-Transformer                   & 0.360               & 0.667                   & 0.457          & -                    & Longformer                 & 2,048                & \cite{jiang2020long}       \\
 & 9 & TKL                               & 0.329              & 0.644                   & -              & -                    & Glove                      & 4,000                & \cite{hofstatter2020local} \\ \bottomrule
\end{tabular}
\end{table*}


%% file: Table/exp_seg_sel2.tex
\begin{table}[]
\caption{
Performance of  segment ranking based on the ground truth of the MS MARCO passage-ranking dataset.
}
\begin{tabular}{lcccc}
\toprule
        & & \multicolumn{2}{c}{\textbf{Segment ranking}} & \textbf{Document ranking} \\ \hline
             & & Train                  & Dev                    & Dev           \\ \hline
             & & P@1 & P@1 & MRR \\ \hline
\multicolumn{2}{l}{Random}       & 0.109                  & 0.085                  & -                \\
\multicolumn{2}{l}{BERT (Max P)} & 0.465                  & 0.391                  & 0.349            \\ \hline 
\multirow{4}{*}{\ProposedMethod} & 
Iter 1  & 0.412                  & 0.412                  & 0.371            \\
& Iter 2  & 0.488                  & 0.449                  & 0.368            \\
& Iter 3  & 0.522                  & \textbf{0.491}                  & \textbf{0.388}            \\
& Iter 4  & \textbf{0.525}                  & 0.458                  & 0.364           \\ \bottomrule
\end{tabular}
\label{tab:seg_sel}
\vspace{-6mm}
\end{table}

%% file: 9_conclusion.tex
\section{Conclusion}

We proposed \ProposedMethod, a query-driven segment selector to build training data for transformer-based rankers that are unable to process long sequences.
Our experiments show that the explicit selection of training data could improve the performance. It shows that challenge of scoring long documents can be addressed without expensive attentions or complex hierarchical architectures.


%% file: 10_acknowledgment.tex
\section*{Acknowledgments}
This work was supported in part by the Center for Intelligent Information Retrieval and in part by NSF grant \#IIS-1813662. 